\documentclass[preprint, prb, showpacs]{revtex4}
\usepackage{epsfig}
\begin{document}
\topmargin-1.0cm

\title {
First principles investigation of transition-metal doped group-IV semiconductors: R${_x}$M$_{1-x}$ (R=Cr, Mn, Fe; M=Si, Ge)
}

\author{Hongming Weng}
\author {Jinming Dong}\email[Corresponding author E-mail:]{jdong@nju.edu.cn}
\affiliation {Group of Computational Condensed Matter Physics, National Laboratory of Solid State Microstructures and Department of Physics, Nanjing University, Nanjing 210093, People's Republic of China}

\date{\today}

\begin{abstract}

A number of transition-metal (TM) doped group-IV semiconductors, R$_{x}$M$_{1-x}$ (R=Cr, Mn and Fe; M=Si, Ge), have been studied by the first principles calculations. The obtained results show that antiferromagnetic (AFM) order is energetically more favored than ferromagnetic (FM) order in Cr-doped Ge and Si with $x$=0.03125 and 0.0625. In 6.25\% Fe-doped Ge, FM interaction dominates in all range of the R-R distances while for Fe-doped Ge at 3.125\% and Fe-doped Si at both concentrations of 3.125\% and 6.25\%, only in a short R-R range can the FM states exist. In the Mn-doped case, the RKKY-like mechanism seems to be suitable for the Ge host matrix, while for the Mn-doped Si, the short-range AFM interaction competes with the long-range FM interaction. The different origin of the magnetic orders in these diluted magnetic semiconductors (DMSs) makes the microscopic mechanism of the ferromagnetism in the DMSs more complex and attractive.
\end{abstract}

\maketitle

\section{introduction} \label{introduction}
Diluted magnetic semiconductors (DMSs) have stimulated a great deal of interests because of its potential applications in the spintronics, in which the electron spin becomes another degree of freedom in addition to the usual charge one. Since the discovery of FM order in Mn doped III-V semiconductors such as InAs\cite{Munekata} and GaAs,\cite{boeck,ohno1,ohno2} the transition metal magnetic impurities have been used as spin injector, which are doped into the semiconductor hosts to cause the ferromagnetism. But usually, the Curie temperatures of these materials are far below the room temperature, which, for example, is 35 K for In$_{1-x}$Mn$_{x}$As and 110 K for Ga$_{1-x}$Mn$_{x}$As. And the II-VI semiconductors are less attractive because the superexchange interaction of the doped magnetic ions favors the AFM or spin glass configuration. However, in some DMSs based upon the transition metal oxide, the ferromagnetism have been observed at or even higher than room temperature, such as 280 K for ZnO\cite{ueda} and 400 K for TiO$_{2}$.\cite{matsumoto} On the other hand, in the group-IV semiconductor, Park {\it et al.}\cite{park} reported that Mn doped  Ge (Mn$_{x}$Ge$_{1-x}$) has its Curie temperature up to 116 K and then Cho {\it et al.} improved it to 285 K.\cite{cho} So, if it were possible to make room temperature FM Mn$_{x}$Si$_{1-x}$, the ``spintronic semiconductor'' industry would grow up rapidly based upon now mature Si-based semiconducting technology and associated facilities. 

The room temperature FM DMSs discovered continuously in experiments brings challenge to the theoretical work because their origin of ferromagnetism is still an open question. Though some different mechanisms, such as the Ruderman-Kittel-Kasuya-Yoshida (RKKY) interaction,\cite{matsukura} double exchange,\cite{akai} double resonance,\cite{inoue} the Zener tunneling model,\cite{dietl} and the mean-field theory,\cite{yagi} have been proposed, none could give a conclusive interpretation. The strong $p-d$ exchange interaction intermediated by mobile holes is thought as the origin of ferromagnetism in III-V compound based DMSs,\cite{ohno2} and carrier-induced ferromagnetism with the exchange interaction mediated by electrons was considered to be suitable to the Co-doped anatase TiO$_{2}$ system.\cite{matsumoto} While the first-principles calculation on the Mn-doped Ge made by Park {\it et al.} indicates that the FM order arises from a long-range FM interaction  competing with a short-range AFM one. More detailed study\cite{zhao} by Zhao {\it et al.} shows that Mn$_{x}$Ge$_{1-x}$ is a RKKY-like FM semiconductor. In this paper, the similar study is extended to the Cr (Fe)-doped Ge and Si. Since a Cr (Fe) atom has one electron less (more) than a Mn atom and Si is in the same column as Ge in the Periodic Table, comparison between these DMSs allows to investigate variations of magnetic properties with change of dopants and host semiconductors. In fact, there have been experiments to be done on the Cr- and Fe-doped Ge\cite{choicrfe, kioseoglou} and also the Mn-doped Si.\cite{hwa} In the follwoing Section II, we will introduce our calculation methods, and our numerical results are shown in the Section III, from which some discussions and conclusions are made.

\section{Calculation Method} \label{Calculation Method}
The software package VASP (Ref. 18) has been used in our calculations, which is based on a total energy pseudo-potential plane-wave method within the local spin density approximation (LSDA). In the calculation, the interaction between ions and electrons is described by the projector-augmented wave method in the generalized gradient approximations (GGA).\cite{paw} The initial crystal structures of R$_{x}$M$_{1-x}$ (R=Cr, Mn and Fe; M=Ge, Si) are taken as $2a\times2a\times2a$ supercell for $x$=0.03125 and $2a\times2a\times1a$ for $x$=0.0625 with two M atoms replaced by two R atoms, among which, the first is at the origin, and the other is put on a lattice position farther away from the origin indicated by three digits following $N$ to represent its $(x,y,z)$ coordinate in units of $a/4$.\cite{zhao} And the lattice constant $a$ is taken as that of pure Ge and Si, i.e., 5.658 $\AA$ and 5.431 $\AA$, respectively. In the spin optimization, the initial spin configuration in the AFM state is taken as 5 net spin on one R atom, and -5 on the other one, and in the FM state, 5 net spins are chosen for both R atoms. The same ground state is reached while increasing the value to 8 and 10. An energy of 350 eV is used for the plane wave cutoff, and when the energy cutoff is increased to 550 eV, the total energy difference between the AFM state and the corresponding FM state changes no more than 0.05 meV/R. For the Brillouin Zone sampling, we take the same k mesh as that in Ref. 14 for all the cases. 

\section{Results and Discussion} \label{Results and Discussion}

The calculated FM total energies and the energy difference between AFM and FM states of the Cr-doped Ge and Si are presented in Tables. I and II for $x$=0.03125 and 0.0625, respectively. Also shown are the averaged magnetic moments on the Cr atom. Obviously, for the Cr-doped Ge at both concentrations, the AFM order is energetically more favored than the FM order, which is consistent with the experimental measurement and theoretical calculation in Ref. 15. The more holes caused by Cr than Mn in Ge host matrix unexpectedly do not enhance the ferromagnetism in this system. And in the Si host matrix, independent of the doping concentration, only for the $N220$ configuration the FM order is a little favorable while the others tend to taking the AFM order. But in general the energy differences between AFM and FM decreases when $x$ changes from 0.0625 to 0.03125, indicating that perhaps at lower enough $x$, the FM state would be lower than AFM in energy, which still needs more experiments to be confirmed.  

Tables. III and IV give the results of Mn-doped Ge and Si with $x$=0.03125 and 0.0625, respectively. It is clear that our results of Mn$_{x}$Ge$_{1-x}$ are the same as those in Ref. 14, indicating that it is a RKKY-like FM semiconductor. However, it is totally different for Mn$_{x}$Si$_{1-x}$ although both Ge and Si are in the same column in the Periodic Table. Independent of the doping concentration, the Mn$_{x}$Si$_{1-x}$ tends to be FM even when the Mn-Mn distance is as long as 9.41 $\AA$ in the $N444$ case except that the $N111$ shows AFM order. The AFM interaction between Mn ions exists only in a short range (the nearest-neighbor, about 2.35 $\AA$), which competes with the long range (all beyond the nearest neighbor) FM interaction. This mechanism was once thought as the origin of FM order in Mn$_{x}$Ge$_{1-x}$ system,\cite{park} which now seems to be responsible for the FM order in Mn doped Si. So, it is plausible that the FM order would be easier in Si host matrix than in Ge for Mn-doping. When $x$=0.03125, the lowest FM energy is found in $N440$ configuration for Ge and in $N220$ for Si, and the corresponding energy differences between AFM and FM states ($E_{AFM}-E_{FM}$) are 103.92 and 74.88 meV/Mn in these two condigurations. Increasing $x$ to 0.0625, the lowest FM energy configuration becomes $N220$ for both Ge and Si, and the corresponding $E_{AFM}-E_{FM}$ increases to 122.2 and 84.61 meV/Mn, respectively. So, in the same doping concentration, the Curie temperature of Mn$_{x}$Ge$_{1-x}$ will be higher than that of Mn$_{x}$Si$_{1-x}$, and in both host matrixes, the Curie temperature will increase with Mn doping concentration.\cite{zhao, stroppa, cho, hwa} 

The results of 3.125\% and 6.25\% Fe-doped the two group-IV semiconductors are listed in Tables. V and VI, respectively. For both concentrations of Fe-doped Ge and Si, the lowest energy state are all the $N111$ configuration with FM order, and all other configurations have much higher energies than the N111. So, there would be ferromagnetism in Fe-doped group-IV semiconductors, and the recent experimental result shows that Fe$_{x}$Ge$_{1-x}$\cite{choicrfe} is a $n$-type FM semiconductor with Curie temperature as high as 233 K while the ferromagnetism in Fe-doped Si still needs further experimental observations. It is found from Table. V, for Ge host matrix, all other configurations, except the $N111$ and $N220$, converge to non-magnetic state even from the initial FM one, and their energies are lower than that of AFM state. In contrast, the $N220$ favours the AFM state rather than the non-magnetic. However, for Si host matrix, all configurations, except only $N111$, favour the non-magnetic, being independent of the initial magnetic state. Comparing these results with those at 6.25\% in Table. VI, we found that the higher Fe concentration will enhance the ferromagnetism. For examples, the FM state in all Fe$_{0.0625}$Ge$_{0.9375}$ configurations are favorable in energy than the AFM. So in the Fe-doping cases, the FM interaction is a very short-range one, and would become longer-range by increasing the doping concentration. 

It is generally thought that the $d$ orbital on the R atom is much more hybridized with $p$ orbital of Si than with that of Ge, which would cause a smaller local magnetic moment on R in Si matrix than that in Ge one.\cite{stroppa} From our calculated magnetic moments on R atoms listed in the above mentioned tables it can be seen that at the same concentrations, in general, the magnetic moment on R in Ge host matrix is a little larger than that in Si one. We have also checked that if the lattice constant of Ge is compressed to that of Si, i. e., increasing the mixing of the $p$ orbital of Ge with the $d$ orbital of R atom, the magnetic moment on R atom would decreas in all cases. Especially in the case of 6.25\% Fe-doping, the decrease of Ge lattice constant will cause the Fe atom to be non-magnetic as like that in Fe$_{0.0625}$Si$_{0.9375}$ listed in the right part of Table. VI. And vice versa, enlarging Si lattice constant to that of Ge causes the $N111$ and $N220$ of the Fe-doped Si to be in FM order although $N400$ and $N440$ reamin to be in the non-maganetic. So, compressing (enlarging) lattice would decrease (enhance) the FM order of the TM-doped group-IV semiconductor, which is consistent with the theoretical calculation on the Mn-doped diamond, another group-IV element with the same crystal structure, in which the absence of ferromagnetism had been predicted.\cite{erwin}

To further investigate the transition metals (TMs) doped group-IV semiconductors, we have also studied the electronic structures of these systems, and focused our attention to the lowest energy FM states in all the configurations. The obtained total DOS and $3d$-partial DOS on R atoms in our concerned systems are shown in Figs. 1 and 2 for $x$=0.03125 and 0.0625, respectively. Since Cr (Fe) has one electron less (more) than Mn atom, a careful analysis of the total DOS of Cr$_{x}$M$_{1-x}$ (Fe$_{x}$M$_{1-x}$) in both figures shows that the Fermi level is a little down-(up-) shifted compared with that of Mn$_{x}$M$_{1-x}$. Moreover, from the $3d$-partial DOS and the magnetic moment of R atoms listed in Table. I to VI, we can deduce that one absent (redundant) electron in Cr (Fe) compared to Mn is in spin up (down) state, which can also be seen from the emergency (narrower) of the gap in spin up (down) channel in the case of Cr (Fe) doping compared to that of the Mn doping. These changes of total DOS around the E$_{F}$ in the spin channels will lead to different spin polarized conductivity in these systems. As shown obviously in Figs. 1 and 2, the Cr- (Fe-) doped Ge and Si system is nearly semiconducting (metal) in the total DOS structure, which is consistent with its tendency to be in AFM (FM) groud state because the decreasing (increasing) of the carriers mobility would suppress (enhance) the FM order in the system. The $3d$ partial DOS of the R atoms are mostly distributed around E$_{F}$ in the range from -3 to +2 eV. And in the spin up channel, the main peaks of Cr in its valence band are centered at about -1 eV while those of Mn and Fe are at -2.3 eV,\cite{stroppa} which are mostly contributed by $e_{g}^{\uparrow}$ and $t_{2g}^{\uparrow}$. But for the spin down part, the occupied peaks of Cr and Fe are mainly located at -0.5 eV and that of Mn is at about -1 eV, all of which are mostly composed of $t_{2g}^{\downarrow}$, whereas the un-occupied state around 0.5 eV for Mn and Fe and that at 1.2 eV for Cr have $e_{g}^{\downarrow}$ character. So it is known from the above results that in general the crystal field splitting $\Delta_{e_{g}-t_{2g}}$ is about 0.5, 1.3 and 1.8 eV and the exchange splitting of $e_{g}$ orbital is about 2.2, 2.8 and 2.8 eV for Cr, Mn and Fe, respectively. Clearly the exchange splitting energy is larger than the crystal field splitting energy in these systems, leading to their quite strong spin ordering and little changes in host's geometrical structure when doped with TMs.\cite{cho,hwa} And above all, we can see that for Mn in Ge (Si), the local magnetic moment is contributed by three electrons in $t_{2g}^{\uparrow}$ orbital and the $e_{g}^{\uparrow}$ orbital is occupied by one itinerant electron. Since a small crystal field splitting, both of them are hybridized with the $p$ orbital of Ge (Si) strongly, which enhances the carriers mobility and reduces the magnetic moment on Mn atom. For the Cr, lack of the one $e_{g}^{\uparrow}$ electron makes this system semiconducting and prefer more to stay in the the AFM order. While in the Fe-doped cases, the one more electon occupies the $t_{2g}^{\downarrow}$ orbital, near the Fermi level, hybridizing with the $p$ eletrons, which makes the gap in the spin-donw channel narrower or disapper, causing the magnetic moment on Fe is smaller than that on Mn.  

According to Anderson's $s-d$ mixing model,\cite{anderson} the local magnetic moment on the transition metal atom doped into the host matrix is determined by three factors: the on-site Coulomb interaction $U$ of $d$ electrons, the mixing of $d$ orbital with the delocalized orbital $s$ of the host and the energy difference between the $d$ orbital and the Fermi level of the system. So, we have also done the LSDA+$U$ calculations within VASP implement, which show that the on-site $U$ will enhance the local magnetic moment on R, especially for the 6.25\% Fe-doped Si. A $U$=3.0 eV would cause a magnetic moment as large as 3.02 $\mu_{B}$ on Fe atom in $N400$ configuration and the corresponding FM state is energically lower than the AFM one by about 31.67 meV/Fe comparing with the non-magnetic state without $U$. And the total density of states (DOS) of its FM state is nearly half-metal, which is consistent with the nearly integer magnetic moment on Fe. The same calculation of Mn$_{0.0625}$Ge$_{0.9375}$ in $N220$ configuration shows that the system remains to be a half-metal\cite{park2} though the magnetic moment on Mn is increased to 3.77 $\mu_{B}$. The reason why half-metal feature is kept even including $U$ is due to the strong $s-d$ mixing in these TM-doped group-IV semdiconductors. Further, the LSDA+$U$ calculation of 6.25\% Cr-doped Si in $N220$ and $N400$ configuration shows that the AFM state is still more stable than the FM one in energy.

\section{Conclusion} \label{Conclusion}
In summary, we have studied the transition metal doped group-IV semiconductors R$_{x}$M$_{1-x}$ (R=Cr, Mn and Fe; M=Si, Ge) by the first principle calculations in LSDA and GGA formalism. The obtained results show that there exist different ground states for different R elements. The higher hole concentration induced by Cr than Mn seems to disfavour the FM ordering. On the contrary, Fe-doping makes the group-IV semiconductors more FM than Mn-doping. It is also found that enlarging the lattice constant would decrease the $p-d$ mixing and be benefit to the FM order, which perhaps is another way to get higher Curie temperature magnetic group-IV semiconductor in experiments. These systems seem to have different mechanism of FM order although their R elements are in the same row and both Si and Ge are in the same column in the Periodic Table, which still needs much more experimental and theoretical efforts to make clear.

\begin{acknowledgments}
The authors thank support to this work from National Science Foundation under Grant No. 90103038. The calculations in this work have been done on the SGI origin 2000 and 3800 Computers.
\end{acknowledgments}


\clearpage
\begin{table}
\caption{Total energy of the FM phase ($E_{FM}$) in different configurations counted relatively to that with lowest FM energy, and the energy difference between AFM and FM state ($E_{AFM}-E_{FM}$) calculated for Cr-doped Ge and Si at $x$=0.03125. Also shown are the average magnetic moments on each Cr atom calculated from initial FM ($M_{FM}$) and AFM ($M_{AFM}$) configuration, respectively. Energies are all in unit of meV/Cr and the magnetic moment are in the unit of $\mu_{B}$/Cr.}
\label{tab1}
\begin{tabular}{c|c|c|c|c|c|c||c|c|c|c}
\hline
\hline
\multicolumn{3}{c|}{} & 
\multicolumn{4}{c||}{Cr$_{0.03125}$Ge$_{0.96875}$} & 
\multicolumn{4}{c}{Cr$_{0.03125}$Si$_{0.96875}$}   \\
\hline
\multicolumn{3}{c|}{System} & 
$E_{FM}$& 
$E_{AFM}$-$E_{FM} $ & 
$M_{FM}$ &
$M_{AFM}$  &
$E_{FM}$& 
$E_{AFM}$-$E_{FM}$ &
$M_{FM}$ &
$M_{AFM}$  \\
\hline
\multicolumn{3}{c|} {$N111$} & 397.61 & -387.295 & 2.10 & $\pm$2.78 & 237.265 & -347.695 & 1.82 & $\pm$2.32 \\    
\multicolumn{3}{c|} {$N220$} & 34.045 & -20.65   & 2.54 & $\pm$2.68 & 14.585	&   11.42  & 2.22 & $\pm$2.27 \\  
\multicolumn{3}{c|} {$N400$} & 0      & -18.5    & 2.54 & $\pm$2.57 &   0     &    -10.8 & 2.22 & $\pm$2.22 \\
\multicolumn{3}{c|} {$N224$} & 21.255 & -28.375  & 2.53 & $\pm$2.61 & 22.795	& -13.425  & 2.20 & $\pm$2.23 \\  
\multicolumn{3}{c|} {$N440$} & 39.365 & -61.555  & 2.48 & $\pm$2.72 & 17.555	&   -23.18 & 2.18 & $\pm$2.29 \\  
\multicolumn{3}{c|} {$N444$} & 28.51  & -31.275  & 2.51 & $\pm$2.61 & 15.61   &  -10.98  & 2.20 & $\pm$2.24 \\  
\hline
\hline
\end{tabular}
\end{table}

\begin{table}
\caption{The same as that in Table. I except at $x$=0.0625}
\label{tab2}
\begin{tabular}{c|c|c|c|c|c|c||c|c|c|c}
\hline
\hline
\multicolumn{3}{c|}{} & 
\multicolumn{4}{c||}{Cr$_{0.0625}$Ge$_{0.9375}$} & 
\multicolumn{4}{c}{Cr$_{0.0625}$Si$_{0.9375}$} \\
\hline
\multicolumn{3}{c|}{System} & 
$E_{FM}$& 
$E_{AFM}$-$E_{FM} $ & 
$M_{FM}$ &
$M_{AFM}$  &
$E_{FM}$& 
$E_{AFM}$-$E_{FM}$ &
$M_{FM}$ &
$M_{AFM}$  \\
\hline
\multicolumn{3}{c|} {$N111$} & 370.58 & -381.27 & 2.09 & $\pm$2.78 & 237.185 & -351.02 & 1.84 & $\pm$2.33 \\ 
\multicolumn{3}{c|} {$N220$} & 10.55  & -36.54  & 2.50 & $\pm$2.69 & 8.72    & 5.825   & 2.21 & $\pm$2.27 \\      
\multicolumn{3}{c|} {$N400$} & 0      & -69.78  & 2.45 & $\pm$2.69 & 0       & -32.165 & 2.17 & $\pm$2.29 \\      
\multicolumn{3}{c|} {$N440$} & 12.09  & -67.72  & 2.44 & $\pm$2.72 & 6.005   & -25.025 & 2.17 & $\pm$2.29 \\       
\hline
\hline
\end{tabular}
\end{table}

\begin{table}
\caption{The same as that in Table. I, except for the Mn-doped Ge and Si.}
\label{tab3}
\begin{tabular}{c|c|c|c|c|c|c||c|c|c|c}
\hline
\hline
\multicolumn{3}{c|}{} & 
\multicolumn{4}{c||}{Mn$_{0.03125}$Ge$_{0.96875}$} & 
\multicolumn{4}{c}{Mn$_{0.03125}$Si$_{0.96875}$}   \\
\hline
\multicolumn{3}{c|}{System} & 
$E_{FM}$& 
$E_{AFM}$-$E_{FM} $ & 
$M_{FM}$ &
$M_{AFM}$  &
$E_{FM}$& 
$E_{AFM}$-$E_{FM}$ &
$M_{FM}$ &
$M_{AFM}$  \\
\hline
\multicolumn{3}{c|} {$N111$}   &     263.27  &   -257.395 & 3.06 & $\pm$3.20 &   26.18    &    -181.135 & 1.18 & $\pm$2.69 \\        
\multicolumn{3}{c|} {$N220$}   &     10.195  &   76.425   & 3.17 & $\pm$3.22 &   0        &    74.88    & 2.83 & $\pm$2.48 \\        
\multicolumn{3}{c|} {$N400$}   &     60.6    &   -17.525  & 3.16 & $\pm$3.21 &   136.995	&    5.38     & 2.81 & $\pm$2.81 \\
\multicolumn{3}{c|} {$N224$}   &     70.355  &   -3.91    & 3.18 & $\pm$3.23 &   142.61	&    23.735   & 2.82 & $\pm$2.80 \\
\multicolumn{3}{c|} {$N440$}   &     0       &   103.92   & 3.16 & $\pm$3.23 &   22.34	&    125.145  & 2.83 & $\pm$2.59 \\
\multicolumn{3}{c|} {$N444$}   &     94.295  &   -27      & 3.16 & $\pm$3.25 &   153.065	&    15.645   & 2.81 & $\pm$2.83 \\
\hline
\hline
\end{tabular}
\end{table}

\begin{table}
\caption{The same as that in Table. III except at $x$=0.0625.}
\label{tab4}
\begin{tabular}{c|c|c|c|c|c|c||c|c|c|c}
\hline
\hline
\multicolumn{3}{c|}{} & 
\multicolumn{4}{c||}{Mn$_{0.0625}$Ge$_{0.9375}$} & 
\multicolumn{4}{c}{Mn$_{0.0625}$Si$_{0.9375}$}   \\
\hline
\multicolumn{3}{c|}{System} & 
$E_{FM}$& 
$E_{AFM}$-$E_{FM}$ & 
$M_{FM}$ &
$M_{AFM}$  &
$E_{FM}$& 
$E_{AFM}$-$E_{FM}$ &
$M_{FM}$ &
$M_{AFM}$  \\
\hline
\multicolumn{3}{c|} {$N111$}  &   278.975   &   -219.87 & 3.03 & $\pm$3.18 &  158.47  &  -258.525 & 2.40 & $\pm$2.69 \\
\multicolumn{3}{c|} {$N220$}  &   0         &   122.2   & 3.15 & $\pm$3.15 &  0       &  84.61    & 2.80 & $\pm$2.55 \\
\multicolumn{3}{c|} {$N400$}  &   4.93      &   104.88  & 3.14 & $\pm$3.19 &  23.275  &  109.76   & 2.76 & $\pm$2.63 \\
\multicolumn{3}{c|} {$N440$}  &   32.655    &   98.185  & 3.13 & $\pm$3.19 &  42.54   &  108.355  & 2.81 & $\pm$2.65 \\
\hline
\hline
\end{tabular}
\end{table}

\begin{table}
\caption{The same as that in Table. I except for Fe-doped Ge and Si.}
\label{tab5}
\begin{tabular}{c|c|c|c|c|c|c||c|c|c|c}
\hline
\hline
\multicolumn{3}{c|}{} & 
\multicolumn{4}{c||}{Fe$_{0.03125}$Ge$_{0.96875}$} & 
\multicolumn{4}{c} {Fe$_{0.03125}$Si$_{0.96875}$}   \\
\hline
\multicolumn{3}{c|}{System} & 
$E_{FM}$& 
$E_{AFM}$-$E_{FM} $ & 
$M_{FM}$ &
$M_{AFM}$  &
$E_{FM}$& 
$E_{AFM}$-$E_{FM}$ &
$M_{FM}$ &
$M_{AFM}$  \\
\hline
\multicolumn{3}{c|} {$N111$} & 0      & 170.545 & 2.22  & $\pm$2.17 & 0      	& 95.33 & 1.96  & $\pm$1.43  \\
\multicolumn{3}{c|} {$N220$} & 246.62 & -81.595 & 0.001 & $\pm$2.47 & 129.325	& 0.03  & -0.001 & $\pm$0.111 \\
\multicolumn{3}{c|} {$N400$} &184.605 & 43.49  & -0.005 & $\pm$2.13 & 81.725 	& -0.01 & 0.000  & $\pm$0.000 \\
\multicolumn{3}{c|} {$N224$} & 204.575 & 39.6  & -0.011 & $\pm$2.11 & 104.61 	& 0     & 0.000  & $\pm$0.001 \\
\multicolumn{3}{c|} {$N440$} & 198.14  & 7.895   & -0.006 & $\pm$2.62 & 86.355 & 0    & 0.000  & $\pm$0.006 \\
\multicolumn{3}{c|} {$N444$} & 211.885 & 45.69   & -0.004 & $\pm$2.15 & 97.65  & 0    & 0.000  & $\pm$0.000 \\
\hline
\hline
\end{tabular}
\end{table}

\begin{table}
\caption{The same as that in Table. V except at $x$=0.0625.}
\label{tab6}
\begin{tabular}{c|c|c|c|c|c|c||c|c|c|c}
\hline
\hline
\multicolumn{3}{c|}{} & 
\multicolumn{4}{c||}{Fe$_{0.0625}$Ge$_{0.9375}$} & 
\multicolumn{4}{c}{Fe$_{0.0625}$Si$_{0.9375}$} \\
\hline
\multicolumn{3}{c|}{System} & 
$E_{FM}$& 
$E_{AFM}$-$E_{FM}$ & 
$M_{FM}$ &
$M_{AFM}$  &
$E_{FM}$& 
$E_{AFM}$-$E_{FM}$ &
$M_{FM}$ &
$M_{AFM}$  \\
\hline
\multicolumn{3}{c|} {$N111$} & 0       & 156.04 & 2.23 & $\pm$2.22 & 0       & 108.97 & 1.98  & $\pm$1.51 \\
\multicolumn{3}{c|} {$N220$} & 114.445 & 38.485 & 2.45 & $\pm$2.39 & 249.405 & -95.02 & 1.57  & $\pm$0.21 \\
\multicolumn{3}{c|} {$N400$} & 97.825  & 94.645 & 2.50 & $\pm$2.06 & 101.655 & 0.015  & 0.004 & $\pm$0.02 \\
\multicolumn{3}{c|} {$N440$} & 129.52  & 70.22  & 2.55 & $\pm$2.05 & 106.87  & 0.045  & 0.02  & $\pm$0.02 \\
\hline
\hline
\end{tabular}
\end{table}

\clearpage

\begin{center}
{\bf Figure Captions}
\end{center}

Fig. 1. The total DOS (thin line) and projected $3d$ partial DOS (bold line) for the lowest FM configurations at 3.125\% doping concentration. The positive DOS value means that of spin up while negative is for spin down. The vertical dashed line represents the Fermi level.

Fig. 2. The same as Fig. 1 but at 6.25\% doping concentration.


\begin{thebibliography}{20}

\bibitem{Munekata}
H. Munekata, H. Ohno, S. von Molnar, Armin Segm\"uller, L. L. Chang, and L. Esaki, Phys. Rev. Lett. {\bf 63}, 1849 (1989). 

\bibitem{boeck}
J. De Boeck, R. Oesterholt, A. Van Esch, H. Bender, C. Bruynseraede, C. Van Hoof, and G. Borghs, Appl. Phys. Lett. {\bf 68}, 2744 (1996).

\bibitem{ohno1}
H. Ohno, A. Shen, F. Matsukura, A. Oiwa, A. Endo, S. Katsumoto, and Y. Iye, Appl. Phys. Lett. {\bf 69}, 363 (1996).

\bibitem{ohno2}
H. Ohno, Science {\bf 281}, 951 (1998).

\bibitem{ueda} 
K. Ueda, H. Tabata, and T. Kawai, Appl. Phys. Lett. {\bf 79}, 988 (2001).

\bibitem{matsumoto} 
Yuji Matsumoto, Makoto Murakami, Tomoji Shono, Tetsuya Hasegawa, Tomoteru Fukumura, Masashi Kawasaki, Parhat Ahmet, Toyohiro Chikyow, Shin-ya Koshihara, Hideomi Koinuma, Science {\bf 291}, 854 (2001).

\bibitem{park}
Y. D. Park, A. T. Hanbicki, S. C. Erwin, C. S. Hellberg, J. M. Sullivan, J. E. Mattson, T. F. Ambrose, A. Wilson, G. Spanos, B. T. Jonker, Science {\textit 295}, 651 (2002).

\bibitem{cho}
Sunglae Cho, Sungyoul Choi, Soon Cheol Hong, Yunki Kim, John B. Ketterson, Bong-Jun Kim, Y. C. Kim and Jung-Hyun Jung, Phys. Rev. B {\bf 66}, 033303 (2002).

\bibitem{matsukura}
F. Matsukura, H. Ohno, A. Shen, and Y. Sugawara, Phys. Rev. B {\bf 57}, R2037 (1998).

\bibitem{akai}
H. Akai, Phys. Rev. Lett. {\bf 81}, 3002 (1998).

\bibitem{inoue}
J. Inoue, S. Nonoyama, and H. Itoh, Phys. Rev. Lett. {\bf 85}, 4610 (2000).

\bibitem{dietl}
T. Dietl, H. Ohno, F. Matsukura, J. Cibert, and D. Ferrand, Science {\bf 287}, 1019 (2000).

\bibitem{yagi}
M. Yagi, K. Noba, and Y. Kayanuma, J. Lumin. {\bf 94-95}, 523 (2001).

\bibitem{zhao}
Yu-Jun Zhao, Tatsuya Shishidou, and A. J. Freeman, Phys. Rev. Lett. {\bf 90}, 047204 (2003)

\bibitem{kioseoglou}
G. Kioseoglou, A. T. Hanbicki, C. H. Li, S. C. Erwin, R. Goswami, and B. T. Jonker, Appl. Phys. Lett. {\bf 84}, 1725 (2003); 

\bibitem{choicrfe}
Sungyoul Choi {\it et. al.}, Appl. Phys. Lett. {\bf 81}, 3606 (2002); G. Kioseoglou, A. T. Hanbicki, and B. T. Jonker, Appl. Phys. Lett. {\bf 83}, 2716 (2003); Sungyoul Choi {\it et. al.}, J. Appl. Phys. {\bf 93}, 7670 (2003)

\bibitem{hwa}
Hwa-Mok Kim, Nam Mee Kim, Chang Soo Park, Shavkat U. Yuldashev, Tae Won Kang, and Kwan Soo Chung, Chem. Mater. {\bf 15}, 3964 (2003).

\bibitem{vasp}
G. Kresse and J. Hafner, Phys. Rev. B {\bf 47}, 558 (1993); ibid. {\bf 49}, 14 251 (1994); G. Kresse and J. Furthm\"uller, Comput. Mat. Sci. {\bf 6}, 15 (1996); G. Kresse and J. Furthm\"uller, Phys. Rev. B {\bf 54}, 11169 (1996).

\bibitem{paw}
G. Kresse and J. Hafner, J. Phys.: Condens. Matt. {\bf 6}, 8245 (1994); G. Kresse and D. Joubert, Phys. Rev. B {\bf 59}, 1758 (1999). 


\bibitem{stroppa}
A. Stroppa, S. Picozzi, A. Continenza, and A. J. Freeman, Phys. Rev. B {\bf 68}, 155203 (2003)

\bibitem{erwin}
Steven C. Erwin and C. Stephen Hellberg, Phys. Rev. B {\bf 68}, 245206 (2003)

\bibitem{anderson}
P. W. Anderson, Phys. Rev. {\bf 124}, 41 (1961)

\bibitem{park2}
J. H. Park, S. K. Kwon, B. I. Min, Physica B, {\bf 281\&282}, 703 (2002)

\end{thebibliography}
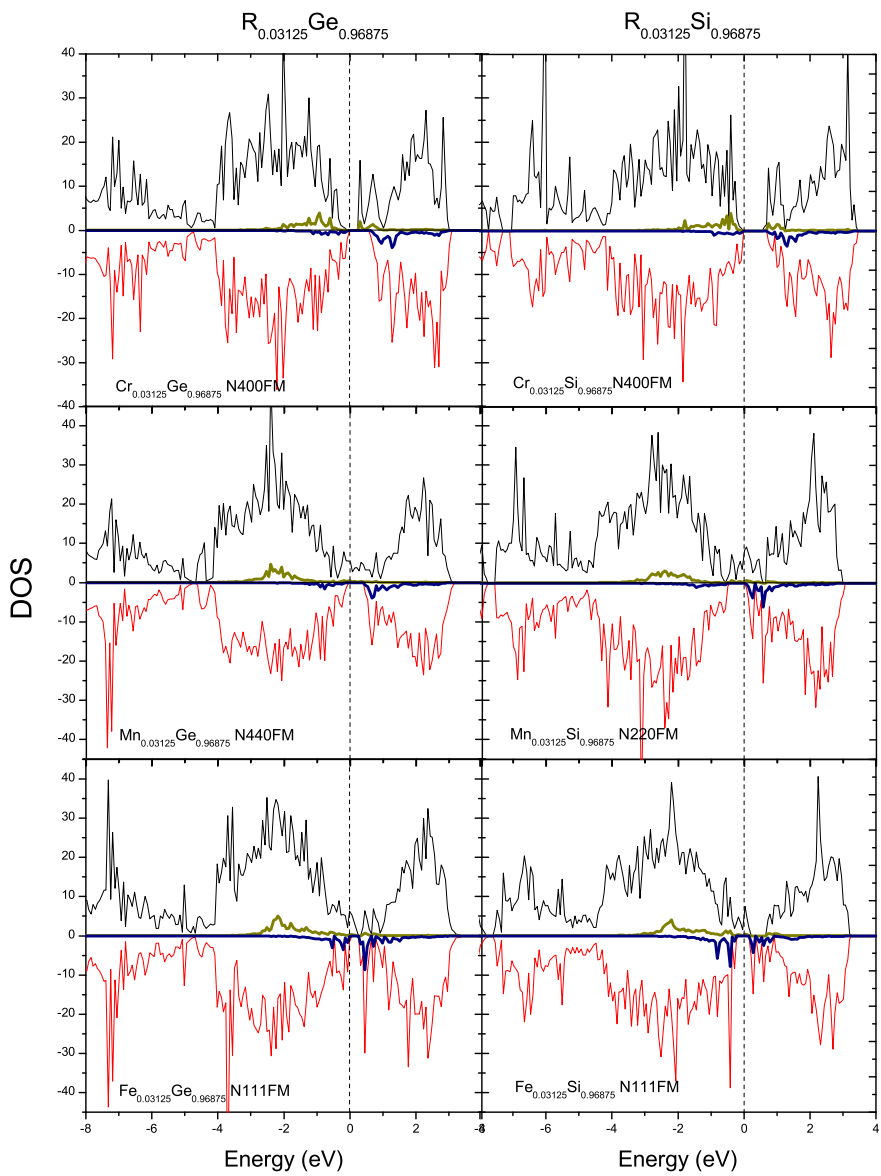
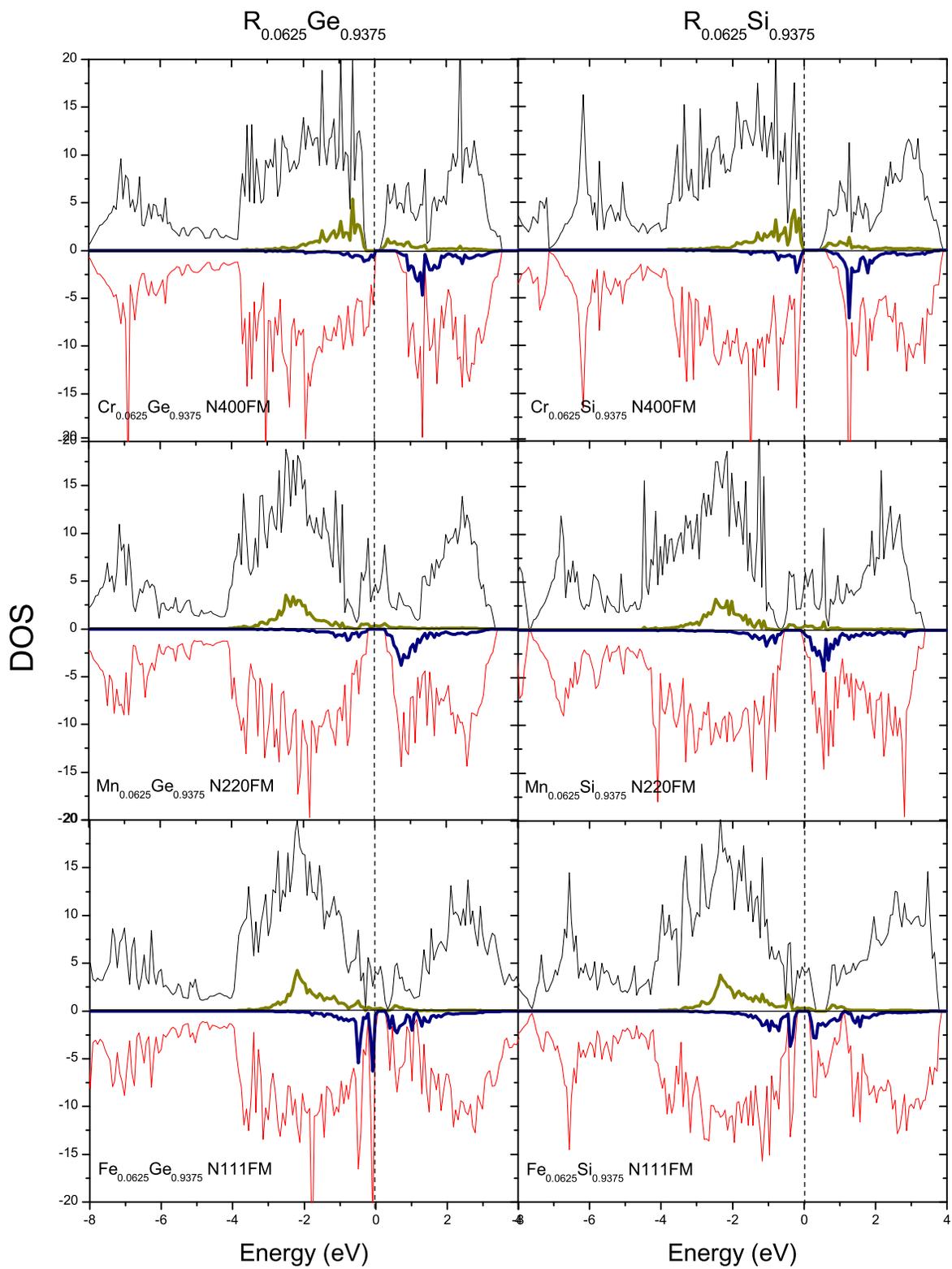
\end{document}